\documentclass[11pt,a4paper]{article}
\pdfoutput=1
\usepackage{graphicx,epsfig,amsmath,jheppub}

\begin{document}

\newcommand{\dd}{\mbox{d}}
\newcommand{\al}{\alpha}
\newcommand{\bt}{\beta}
\newcommand{\lm}{\lambda}
\newcommand{\gm}{\gamma}
\newcommand{\Gm}{\Gamma}
\newcommand{\dl}{\delta}
\newcommand{\Dl}{\Delta}
\newcommand{\ep}{\epsilon}
\newcommand{\vep}{\varepsilon}
\newcommand{\kp}{\kappa}
\newcommand{\Lm}{\Lambda}
\newcommand{\om}{\omega}
\newcommand{\pa}{\partial}
\newcommand{\nn}{\nonumber}
\newcommand{\bea}{\begin{eqnarray}}
\newcommand{\eea}{\end{eqnarray}}
\newcommand{\be}{\begin{equation}}
\newcommand{\ee}{\end{equation}}
\newcommand{\bfm}[1]{\mbox{\boldmath$#1$}}
\newcommand{\bff}[1]{\mbox{\scriptsize\boldmath${#1}$}}
\newcommand{\grtsim}{\;\rlap{\lower 3.5 pt \hbox{$\mathchar \sim$}} \raise 1pt \hbox {$>$}\;}
\newcommand{\lessim}{\;\rlap{\lower 3.5 pt \hbox{$\mathchar \sim$}} \raise 1pt \hbox {$<$}\;}

\newcommand{\si}{\sigma}
\newcommand{\la}{\lambda}
\newcommand{\ga}{\gamma}
\newcommand{\de}{\delta}
\newcommand{\De}{\Delta}
\newcommand{\chill}{\chi_{\lambda \lambda}}
\newcommand{\child}{\chi_{\lambda d}}
\newcommand{\chidl}{\chi_{d \lambda}}
\newcommand{\chidd}{\chi_{dd}}
\newcommand{\ls}{\ln \! \left(\frac{s}{M^2}\right)}
\newcommand{\lsnb}{\ln \! \frac{s}{M^2}}
\newcommand{\lsn}[1]{\ln^{#1} \! \left(\frac{s}{M^2}\right)}
\newcommand{\lsnnb}[1]{\ln^{#1} \! \frac{s}{M^2}}
\newcommand{\lqs}{\ln \! \left(\frac{Q^2}{M^2}\right)}
\newcommand{\lqsnb}{\ln \! \frac{Q^2}{M^2}}
\newcommand{\lqsn}[1]{\ln^{#1} \! \left(\frac{Q^2}{M^2}\right)}
\newcommand{\lqsnnb}[1]{\ln^{#1} \! \frac{Q^2}{M^2}}
\newcommand{\bo}{\beta_0}
\newcommand{\pd}{\partial}
\newcommand{\Real}{\text{Re}}
\newcommand{\Imag}{\text{Im}}
\newcommand{\tr}{\text{tr}\!}
\newcommand{\fs}[1]{\not \! {#1} }
\newcommand{\dst}{\underset{s\leftrightarrow t}{\Delta}}

\newcommand{\epl}{e^+}
\newcommand{\emi}{e^-}
\newcommand{\OO}{\mathcal{O}}
\newcommand{\Amp}{{\cal{A}}}
\newcommand{\Ampn}[1]{{\cal{A}}^{\left({#1}\right)}}
\newcommand{\FormF}{{\cal{F}}}
\newcommand{\Formn}[1]{{\cal{F}}^{\left({#1}\right)}}
\newcommand{\redAmp}{\tilde{\Amp}}
\newcommand{\redAmpn}[1]{\tilde{\Amp}^{\left({#1}\right)}}
\newcommand{\expn}[2]{#1^{\left({#2}\right)}}
\newcommand{\alfp}{\frac{\al}{4\pi}}

\title{Two-loop electroweak corrections to high energy
       large-angle Bhabha scattering}

\author[1,2,3]{A.A.~Penin} 
\author[1]{and G.~Ryan} 

\affiliation{Department of Physics, University Of Alberta, Edmonton, AB T6G 2J1, Canada}
\affiliation{Institut f{\"u}r Theoretische Teilchenphysik, KIT,  76128 Karlsruhe, Germany}
\affiliation{Institute for Nuclear Research of Russian Academy of Sciences, 117312 Moscow, Russia}

\emailAdd{penin@ualberta.ca}
\emailAdd{gsryan@ualberta.ca}

\abstract{
We  compute the dominant  logarithmically enhanced  two-loop electroweak corrections to 
the electron-positron scattering differential cross section in the high
energy limit for large scattering angles.   Depending on the scattering  angle
and energy the two-loop corrections may exceed one percent and are important  for
the high-precision luminosity spectrum determination at a future linear
collider.
}


\maketitle
\flushbottom

\section{Introduction}
\label{int}
High energy electron-positron or {\it Bhabha} scattering \cite{Bha} is among the
most important and carefully studied processes in particle physics. It provides
a very efficient tool for luminosity determination at electron-positron
colliders and thus mediates the process of extracting physical information from
the raw experimental data \cite{Jad,MNP,MCWG}. In particular it is crucial for the
high precision physics program at a future International Linear Collider (ILC).
The small-angle Bhabha scattering is particularly effective as a luminosity
monitor while the  large-angle scattering is suggested  as a tool to disentangle
the luminosity spectrum at the ILC \cite{Too,Heu}. Bhabha scattering involves
stable charged leptons both in the initial and the final states and, therefore,
it can be measured experimentally with very high precision. On the theory side
it can be reliably computed order-by-order in perturbation theory. These
properties make Bhabha scattering an ideal ``standard candle'' for
electron-positron colliders. Since the theoretical accuracy  directly affects
the luminosity determination, remarkable efforts have been devoted to the study of
the radiative corrections. The one-loop corrections have been known in the full
electroweak theory  for several decades \cite{Bhabha1loop}. Recently the calculation
of the two-loop QED corrections has been completed
\cite{Fad1,Arb,Jad1,BDG,Bas,Bon,Pen,BonFer,BecMel,BFP,Act,KuhUcc}. This result
is sufficient for the  small-angle or low-energy process. For the
large-angle scattering above the electroweak scale  the electroweak corrections
become equally important and must be taken into account to match the designed 
experimental  accuracy of a few permill  \cite{TESLA}. The full evaluation
of the electroweak two-loop corrections remains beyond the reach of available
perturbative techniques. However, once the center-of-mass energy $\sqrt{s}$  is
far larger than the electroweak gauge boson mass $M_{W,Z}$,  the cross section
receives virtual corrections enhanced by powers of  ``Sudakov'' logarithm
$\ln\bigl({s/M_{W,Z}^2}\bigr)$ which dominate the two-loop corrections. Over the
last few years the electroweak  Sudakov logarithms have been studied within
different approaches and the corresponding  high-order corrections to various
processes have been obtained
\cite{CiaCom,Fad,KPS,KMPS,Bec,DenPoz,DMP,FKPS,JKPS0,JKPS,Man0,Man,KMP,KMPU}.

In this paper we generalize the next-to-next-to-next-to-leading (N$^3$LL) result
of Ref.~\cite{JKPS} for the  $f\bar f\to f'\bar f'$ four-fermion annihilation
process to  Bhabha scattering.    The result includes all the two-loop
logarithmically enhanced corrections. The quartic, cubic and quadratic
logarithms are obtained in full $SU_L(2)\times U_Y(1)$  electroweak standard
model, while the linear logarithmic term is computed in a simplified $SU_L(2)$
theory sufficient for the practical applications. In the next section we outline
the approach and present the analytical one- and two-loop results for the
differential cross section in an $SU_L(2)$ model. In Section~\ref{ew} we
present the numerical result for the cross section in full electroweak theory.
The main difference in respect to the analysis  given in  Ref.~\cite{JKPS}  is
in the presence of both annihilation and scattering contributions to the Bhabha cross section and
in using the electron mass as a collinear regulator for QED radiative
corrections.  

\section{High-energy asymptotic of radiative corrections in $SU_L(2)$ model}
\label{su2}
We study the electron-positron scattering in the high-energy and wide-angle
limit when all the kinematical invariants are of the same order and  far larger
than the gauge boson mass, $|s|\sim |t| \sim |u| \gg M_{Z,W}^2$. In this limit
the asymptotic energy dependence of the field amplitudes is dominated by {\em
Sudakov} logarithms  \cite{Sud,Jac} and  governed by the evolution equations
\cite{Mue,Col,Sen,Ste}. The method of the evolution equations in the context of
the electroweak corrections is described in detail for  the fermion pair
production in Ref.~\cite{JKPS}. The evolution equations for the scattering
(annihilation) amplitude are linear which results in the exponentiation of the
Sudakov logarithms. To outline the general structure of the corrections let us
for a moment neglect the hypercharge interaction and consider a spontaneously
broken $SU_L(2)$ gauge model with the gauge boson of mass $M$ and twelve
massless left-handed fermion doublets. This model retains the main features of
the massive gauge boson sector of the Standard Model. In this case the result
can be presented in a simple analytical form and constitutes the basis for the
further extension to the full electroweak theory.  The amplitude can be
factorized into the following form
\be
{\cal A}={ig^2(Q^2)\over Q^2}{\cal F}^2\bfm{\tilde{\cal A}}\,,
\label{defa}
\ee
where $g(\mu^2)$ is the  $SU_L(2)$ coupling renormalized in  $\overline{\rm MS}$
scheme, $Q$ is the Euclidean momentum transfer $Q^2=-s$ for annihilation and  
$Q^2=-t$ for the scattering process, ${\cal F}$ is the electron vector
form factor, and the reduced amplitude $\bfm{\tilde{\cal A}}$ is a vector in an
isospin and chiral basis. The solution of the evolution equations read
\be
{\cal F}=F_0(\al(M^2))\exp \left\{\int_{M^2}^{Q^2}{\dd x\over x}
\left[\int_{M^2}^{x}{\dd x'\over x'}\gm(\al(x'))+\zeta(\al(x))
+\xi(\al(M^2))\right]\right\} \,,
\label{evolsolf}
\ee
\be
\bfm{\tilde{\cal A}}={\rm P}\!\exp{\left[\int_{M^2}^{Q^2}
{\dd x\over x}\chi(\al(x))\right]}\bfm{\cal A}_{0}(\al(M^2))\,,
\label{evolsola}
\ee
where  the perturbative functions $F_0(\al)$, $\gm(\al)$ {\it etc.} are given by a series
in the coupling constant $\alpha(\mu^2)=g^2(\mu^2)/(4\pi)$. By calculating
the functions entering the evolution equations order by order in $\al$ one gets
the logarithmic approximations for the amplitude.  In the $SU_L(2)$ model with
the degenerate Higgs and gauge boson mass the parameters of the
Eqs.~(\ref{evolsolf},\ref{evolsola}) are known to the N$^3$LL approximation
which includes all the two-loop logarithmic corrections \cite{JKPS}. By analytical
continuation and crossing symmetry it is straightforward to find the
electron-positron scattering and annihilation amplitudes for the physical values
of the Mandelstam variables and the corresponding contributions to the cross
section.  

\subsection{Differential cross section}
The perturbative series for the Bhabha cross section is of the
following form
\begin{align}
\frac{\dd \si}{\dd \Omega} &= \frac{\al^2(s)}{64s}\frac{(1-x)^2}{x^2}
\left[ x^2\left(1+\alfp\expn{\dd \si_S}{1}+\frac{\al^2}{16\pi^2}\expn{\dd \si_S}{2}\right)
+\left(1+\alfp \expn{\dd \si_T}{1}+\frac{\al^2}{16\pi^2}\expn{\dd \si_T}{2}\right)\right.
\nonumber \\
&-\left. 2x \left(1+\alfp\expn{\dd \si_{ST}}{1}+\frac{\al^2}{16\pi^2}
\expn{\dd \si_{ST}}{2}\right) \right] +{\cal O}(\al^3)\,.
\label{su2series}
\end{align}
where the variable $x=(1-\cos{\theta})/2$ is a function of the scattering angle
$\theta$, $\al\equiv \al(M^2)$, and  the annihilation $S$, scattering $T$, and
interference $ST$ contributions are given separately. The coupling constant in
the Born approximation is normalized at $\mu^2=s$ and can be expanded as follows
\be
\al(s)=\al\left[1-{\al\over 4\pi}\beta_0\ls+\left({\al\over 4\pi}\right)^2
\left(\beta^2_0\lsn{2}-\beta_1\ls\right)+{\cal O}(\al^3)\right]\,,
\ee
with $\beta_0=19/6$ and  $\beta_1=-35/6$. The corrections can be further split
according to the power of the large logarithms 
\begin{align}
	\expn{\dd \si}{1} &= \expn{\dd \si_2}{1} \lsn{2} + \expn{\dd \si_1}{1} \ls +  \expn{\dd \si_0}{1} \ , \nonumber  \\
	\expn{\dd \si}{2} &= \expn{\dd \si_4}{2} \lsn{4} + \expn{\dd \si_3}{2} \lsn{3}+\expn{\dd \si_2}{2} \lsn{2} + 
	\expn{\dd \si_1}{2} \ls  +  {\cal O}(1)\,,
\end{align}
where $\expn{\dd \si_i}{j}$ are functions of $x$,  $\expn{\dd \si_2}{1}$
corresponds to the one-loop leading logarithmic (LL) correction, $\expn{\dd
\si_1}{1}$ corresponds to the one-loop next-to-leading logarithmic (NLL)
correction, and so on. The high-energy asymptotic of the  one-loop corrections reads
\begin{align}
	\expn{\dd \si_{S\ 2}}{1} &=  -3  \ , \nonumber \\
	\expn{\dd \si_{S\ 1}}{1} &=  9 + 2 \ln(1-x) -10 \ln (x) \ , \nonumber \\
	\expn{\dd \si_{S\ 0}}{1} &=  \frac{235}{18} - 3\pi^2 + \frac{5 \ln(x)}{1-x} - \frac{5(1-2x)\ln^2(x)}{2(1-x)^2}\,,
\hspace*{5mm}
\label{1lsus}
\end{align}

\begin{align}
	\expn{\dd \si_{T\ 2}}{1} &=  -3 \ , \nonumber  \\
	\expn{\dd \si_{T\ 1}}{1} &=  9 +2 \ln(1-x) + 2 \ln (x)  \ , \nonumber \\
	\expn{\dd \si_{T\ 0}}{1} &=  \frac{235}{18} + 2 \pi^2 + 2\ln(1-x)\ln(x) \nonumber \\
	&\qquad+\frac{(8+7x)\ln(x)}{3(1-x)} + \frac{5(2-2x+x^2)\ln^2(x)}{2(1-x)^2} \,,
\end{align}

\begin{align}
	\expn{\dd \si_{ST\ 2}}{1} &=  -3 \ , \nonumber  \\
	\expn{\dd \si_{ST\ 1}}{1} &=  9 +2 \ln(1-x) - 4 \ln (x) \ , \nonumber \\
	\expn{\dd \si_{ST\ 0}}{1} &=  \frac{235}{18} - \frac{1}{2}\pi^2 +\ln(1-x)\ln(x) \nonumber \\
	&\qquad +\frac{(23+7x)\ln(x)}{6(1-x)} + \frac{5(1+x^2)\ln^2(x)}{4(1-x)^2} \,.
\hspace*{3mm}
\end{align}
By reexpanding the exponents~(\ref{evolsolf},~\ref{evolsola}) one gets the
two-loop logarithmic corrections which are given by
Eqs.~(\ref{tlsus}-\ref{tlsust}) in the Appendix A.
Eqs.~(\ref{1lsus},~\ref{tlsus}) also give the corrections to the differential
cross section  of the $e^+e^-\to \mu^+\mu^-$ process or, in general, of a
different flavor same-isospin fermion pair production.  For the production of an 
opposite-isospin fermion pair  such as $e^+e^-\to \nu\bar{\nu}$ the one-loop
coefficients read
\begin{align}
	\expn{\dd \si_{S\ 2}}{1} &=  -3  \ , \nonumber \\
	\expn{\dd \si_{S\ 1}}{1} &=  9 -10 \ln(1-x) +2 \ln (x) \ , \nonumber \\
	\expn{\dd \si_{S\ 0}}{1} &=  \frac{235}{18}-3\pi^2 - \frac{ \ln(x)}{1-x} + \frac{(1-2x)\ln^2(x)}{2(1-x)^2}\,,
\hspace*{5mm}
\end{align}
and the two-loop coefficients can be found in Appendix A. 

\subsection{Total annihilation cross section}
The differential cross section of the  pure annihilation process can be
integrated over  solid angle to get the total cross section
$\sigma=\sum_n(\alpha/4\pi)^n\sigma^{(n)}$, where $\sigma^{(0)}=\pi\alpha^2(s)/(48 s)$ is the total
cross section in the Born approximation. The one- and two-loop corrections to
the total cross section are
\bea
\sigma^{(1)}&=&\left[-3\ln^2\left({s\over M^2}\right)
+{80\over 3}\ln\left({s\over M^2}\right)
-\left({85\over 9}+3\pi^2\right)\right]\sigma^{(0)}\,,
\nn \\
\sigma^{(2)}&=&\left[{9\over 2}\ln^4\left({s\over M^2}\right)
-{461\over 6}\ln^3\left({s\over M^2}\right)
+\left({5875\over 18}+{37\pi^2\over 3}\right)\ln^2\left({s\over M^2}\right)\right.
\nn \\
&&\left.+\left(-{16019\over 216}-{1637\over 18}\pi^2-122\zeta(3)+15\sqrt{3}\pi
+26\sqrt{3}{\rm Cl}_2\left({\pi\over 3}\right)\right)\ln\left({s\over M^2}\right)
\right]\sigma^{(0)}\,
\label{sigsu2s}
\eea
for $e^+e^-\to \mu^+\mu^-$, and
\bea
\sigma^{(1)}&=&\left[-3\ln^2\left({s\over M^2}\right)
+{26\over 3}\ln\left({s\over M^2}\right)
+\left({158\over 9}-3\pi^2\right)\right]\sigma^{(0)}\,,
\nn \\
\sigma^{(2)}&=&\left[{9\over 2}\ln^4\left({s\over M^2}\right)
-{137\over 6}\ln^3\left({s\over M^2}\right)
-\left({451\over 9}-{37\pi^2\over 3}\right)\ln^2\left({s\over M^2}\right)\right.
\nn \\
&&\left.+\left({24481\over 216}-\frac{341}{18}\pi^2-122\zeta(3)+15\sqrt{3}\pi
+26\sqrt{3}{\rm Cl}_2\left({\pi\over 3}\right)\right)\ln\left({s\over M^2}\right)
\right]\sigma^{(0)}\,
\label{sigsu2o}
\eea
for $e^+e^-\to \nu\bar{\nu}$ process. Here ${\rm Cl}_2\left({\pi\over
3}\right)=1.014942\ldots$ is the value of the Clausen function and
$\zeta(3)=1.20206\ldots$ is the value of the Riemann zeta-function. Similar
formulae have been derived in Refs.~\cite{KMPS,JKPS0,JKPS} for a model with six
massless left-handed fermion doublets.\footnote{In  Eq.~(3) \cite{JKPS0} and 
Eq.~(44)  \cite{JKPS} in the coefficient of the single logarithmic term
${48049/216}-{1679\pi^2/18}$ should be replaced by ${38221/216}-{1601\pi^2/18}$.
In  Eq.~(4) \cite{JKPS0}  and Eq.~(45) \cite{JKPS} in the coefficient of the
single logarithmic terms ${38005/216}-{383\pi^2/18}$ should be replaced by
${28177/216}-{305\pi^2/18}$.} Our choice of the fermionic content provides more
accurate model of the electroweak theory with three generations of fermions.   
 
\section{Two-loop electroweak corrections to Bhabha scattering}
\label{ew}
In the full electroweak standard model the gauge coupling depends on the
electron chirality and the cross section can be written as follows 
\be
\frac{\dd \si}{\dd \Omega} = \frac{\al^2}{4 s} \sum_{I,J \in \{L,R\}} \left( \dd\si_{IJ}^S 
+ \dd\si_{IJ}^T +2 \ \dd\si_{IJ}^{ST} \right) \,,
\ee
where the index $I$ $(J)$ stands for the initial (final) electron chirality,
the $\overline{MS}$ $SU_L(2)$  coupling constant  $\al= {\al_e}/{\sin^2 \theta_W}$ is a ratio
of  the  QED coupling constant $\al_e$ an the weak mixing angle $\theta_W$ normalized at $\mu=M_W$. In
the Born approximation one has
\be
\begin{array}{lll}
{\dd \si_{LL}^S}^{(0)} = \left(\frac{1}{4} + \frac{t_W^2}{4}\right)^2(1-x)^2 , & {\dd \si_{RR}^S}^{(0)} = t_W^4 (1-x)^2 , &  {\dd \si_{LR}^S}^{(0)} = {\dd \si_{RL}^S}^{(0)} = \frac{t_W^4}{4}x^2 ,  \\
{\dd \si_{LL}^T}^{(0)} = \left(\frac{1}{4} + \frac{t_W^2}{4}\right)^2\frac{(1-x)^2}{x^2}, & {\dd \si_{RR}^T}^{(0)} = t_W^4 \frac{(1-x)^2}{x^2}, &  {\dd \si_{LR}^T}^{(0)} = {\dd \si_{RL}^T}^{(0)} = \frac{t_W^4}{4}\frac{1}{x^2}, \\
{\dd \si_{LL}^{ST}}^{(0)} =  -\left(\frac{1}{4} + \frac{t_W^2}{4}\right)^2 \frac{(1-x)^2}{x}, &{\dd \si_{RR}^{ST}}^{(0)} = -t_W^4 \frac{(1-x)^2}{x} ,& \dd \si_{LR}^{ST} = \dd \si_{RL}^{ST} = 0\,,
\end{array}
\ee
where $t_W^2\equiv \tan^2{\theta_W}$. The total electroweak Born cross section
of Bhabha scattering then reads
\be
\frac{\dd\si^\text{Born}}{\dd \Omega} = \frac{\al^2}{64\:s \: x^2} 
\Big((1-x)^4 (1+t_W^2)^2+8 t_W^4(3-8x+12x^2-8x^3+3x^4)\Big)\,.
\ee
We replace the hypercharge coupling constant  by $\al \tan^2{\theta_W}$ and 
consider a single-parameter perturbative series for the cross section 
\begin{align}
\frac{\dd \si}{\dd \Omega}& =\frac{\dd\si^\text{Born}}{\dd \Omega}+\frac{\al^2}{4\: s\: x^2}\left[\frac{\al}{4\pi} 
\left(\dd \si_2^{(1)}\ln^2\left(\frac{s}{M_W^2}\right)+\dd \si_1^{(1)} 
\ln\left(\frac{s}{M_W^2}\right)+\dd \si_0^{(1)} \right)\right. \nonumber \\
&\qquad \left.+ \left(\frac{\al}{4\pi}\right)^2\left(\dd \si_4^{(2)} 
\ln^4\left(\frac{s}{M_W^2}\right)+\dd \si_3^{(2)} 
\ln^3\left(\frac{s}{M_W^2}\right)\right.\right.\nonumber \\
&\qquad \left.\left.+\dd \si_2^{(2)} \ln^2\left(\frac{s}{M_W^2}\right)+\dd \si_1^{(2)} 
\ln\left(\frac{s}{M_W^2}\right)+{\cal O}(1)\right)+{\cal O}(\alpha^3)\right] \,.
\label{ewseries}
\end{align}
The  analysis of Sudakov logarithms in electroweak standard model with the
spontaneously broken $SU_L(2)\times U_Y(1)$ gauge group is similar  to the
treatment of the pure $SU_L(2)$ case considered in the previous section. The
main difference is in the presence of the massless photon. The photon
contribution to the virtual corrections results in the QED Sudakov logarithms of
the form $\ln Q^2/\lm^2$ and  $\ln Q^2/m_f^2$, where $m_f$ is a light fermion
mass  and  $\lm$ is the auxiliary photon mass introduced to regulate the
infrared divergences. By factorizing the QED logarithms from the solution of the
full electroweak evolution equations one gets pure electroweak logarithms of the
form $\ln(Q^2/M_{W,Z}^2)$. The QED logarithms in turn form an exponential
factor. In Ref.~\cite{JKPS} this factor has been evaluated to N$^3$LL accuracy
for massless fermions. For our analysis we need only the NNLL approximation. In
the limit  $m_f^2\ll\lm^2\ll M_{W,Z}^2\ll s$,  where $f\ne t$ stands for all the
fermions except the top quark, for the annihilation amplitude it takes the
following form
\bea
{\cal U}&=&\exp{\left\{
{\alpha_e(\lm^2)\over 2\pi}\left[\left(-\left(1
-\left({76\over 27}+{8\over 9}\ln\left({1-x\over x}\right)\right)
{\alpha_e\over \pi}\left(N_g-{1\over 2}\right)\right)
\right)\right.\right.}
\nn\\
& &\times\ln^2\left({s\over \lm^2}\right)+\left(3+
2\ln\left({1-x\over x}\right)\right)\ln\left({s\over\lm^2}\right)
-{8\over 27}{\alpha_e\over \pi}\left(N_g-{1\over 2}\right)
\nn \\
& &
\left.\left.
\times\ln^3\left({s\over\lm^2}\right)\right]
+{\cal O}\left(\al_e^2\ln\left({s\over\lm^2}\right)\right)\right\}\,.
\label{QED}
\eea
Here $N_g=3$ stands for the number of generations and $-1/2$ subtracts the top
quark contribution.  A non-standard limit with $m_f\ll\lm$ has been used to
facilitate the procedure of separating the QED and electroweak  logarithms
since it allows for the calculation with the unbroken theory fields.
The normalization of the QED factor depends on the factorization scheme 
and may include nonlogarithmic corrections in $\al_e$. We chose the normalization 
${\cal U}(s=\lm^2)=1$. The corresponding expression for the scattering amplitude 
may be easily obtained by crossing symmetry and analytical continuation of 
Eq.~(\ref{QED}).

The infrared divergent virtual QED contribution should be combined with the real
photon radiation to get a physically observable infrared finite cross section. In
practice, the massive gauge bosons are supposed to be detected as separate
particles.  Thus it is of little physical sense to treat the hard photons with
energies far larger than $M_{Z,W}$ separately because of gauge symmetry
restoration.  In particular, the radiation of the hard real photons is not of
the Poisson type due to its non-Abelian $SU(2)_L$ component.  Therefore, we
restrict the analysis to semi-inclusive cross sections with the real emission
only of photons with energies far smaller than $M_{Z,W}$, which is of pure QED
nature.  The standard Monte Carlo event generators \cite{BABAYAGA,BHWIDE,MCGPJ}
adopt a different limit on the photon mass $\lm\ll m_f$. Let us now
consider the evolution of the QED factor ~(\ref{QED}) to the small values of
$\lm$. At every threshold $\lm =m_f$ the anomalous dimensions of the
corresponding evolution equation change due to the  decoupling of the massive
virtual fermion and one has to match the two solutions which result in the
logarithms of the form $\ln(s/m_f^2)$ in the exponent. At the lowest threshold 
$\lm =m_e$ there is an additional nonlogarithmic one-loop matching correction
because for   $\lm \ll m_e$ the electron rather than photon mass regulate
collinear divergence. This matching correction can be obtained by comparing the
one-loop corrections to the vector form factor in the two limits \cite{Pen}. The
result for the QED factor, which  is now compatible with the Monte Carlo event 
generators, reads
\bea
{\cal U}&=&\left(1+{\al_e\over\pi}\left(-{1\over 4}
+{\pi^2\over 2}\right)\right)\exp{\left\{
{\alpha_e(m_e^2)\over 2\pi}\left[
-\ln^2\left({s\over m_e^2}\right)
+2\left(\ln\left({s\over m_e^2}\right)-1\right)
\ln\left({\lm^2\over m_e^2}\right)\right.\right.}
\nn\\
& &
+3\ln\left({s\over m_e^2}\right)+2\ln\left({1-x\over x}\right)
\ln\left({s\over\lm^2}\right)+{\alpha_e\over \pi}\bigg[
-{1\over 9}\sum_fQ_f^2\ln^3\bigg({s\over m_f^2}\bigg)
\nn\\
& &
+\left(
{19\over 18}+{1\over 3}\ln\left({1-x\over x}\right)\right)
\sum_fQ_f^2\ln^2\bigg({s\over m_f^2}\bigg)\bigg]\bigg]
+{\cal O}\left(\al_e^2\ln\left({s\over m_e^2}\right)\right)\bigg\}\,.
\label{QED1}
\eea
The sum of the virtual and real QED corrections to the cross section is finite
for $\lm=0$ and gives a factor which depends on the Mandelstam variables, the
fermion masses, and the experimental cuts, but not on $M_{Z,W}$. 

\begin{figure}[t]
\begin{center}
\begin{tabular}{cc}
\includegraphics[height=1.9in]{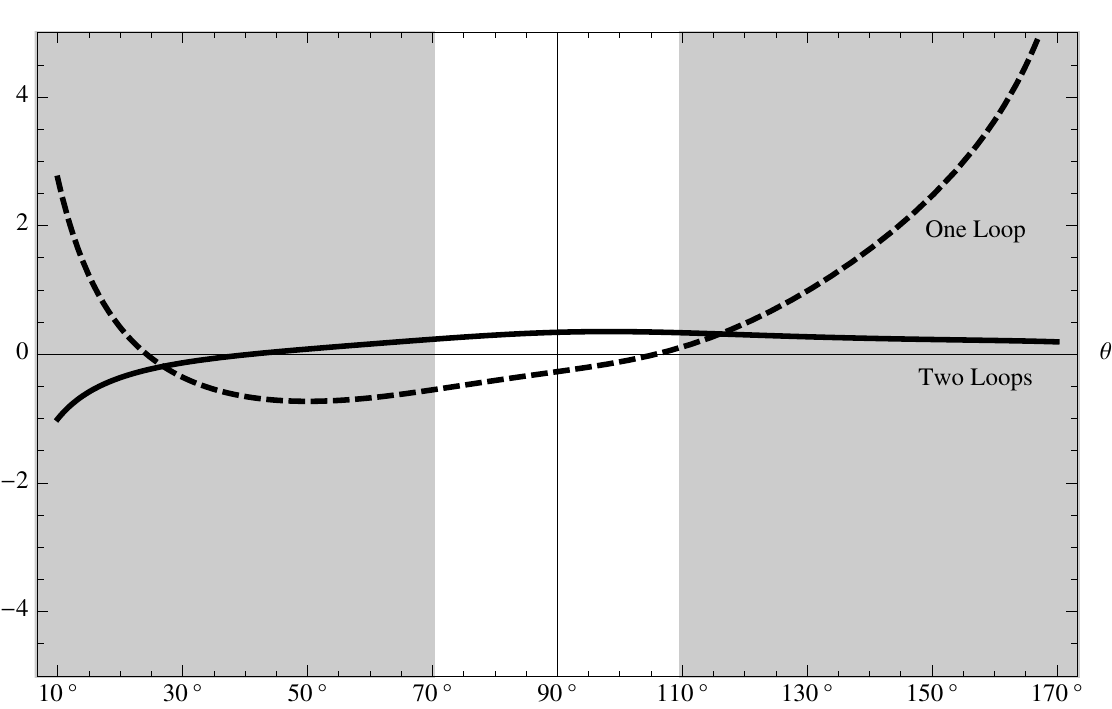} & \includegraphics[height=1.9in]{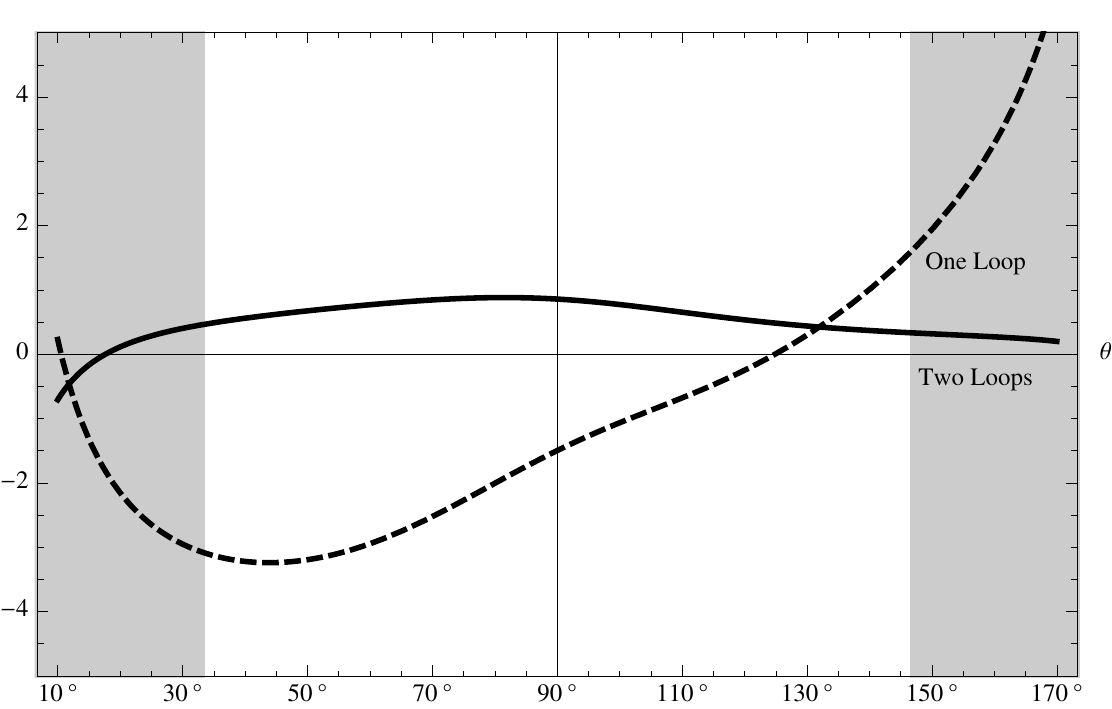} \\
(a) $\sqrt{s} = 500$~GeV & (b)  $\sqrt{s} = 1$~TeV
\end{tabular}
\end{center}
\caption{\label{fig1}   
The one- and two-loop electroweak corrections in \% to the Born cross section
as function of the scattering angle for (a) $\sqrt{s} =500$~GeV
and (b) $\sqrt{s} =1$~TeV. The shaded area corresponds to the region where Sudakov approximation is not reliable.} 
\end{figure}

Thus, in comparison to the $SU_L(2)$  model the evolution equations describing
the electroweak Sudakov logarithms are modified by the presence of the
hypercharge interaction and by the subtraction of the QED logarithms. All the
parameters of the evolution equations are known to the NNLL approximation in the
full electroweak theory including the effect of the gauge boson mass splitting
\cite{KMPS} and the corresponding one- and two-loop corrections can be easily
obtained. We present the result in semi-numerical form for  the following input
values
\be
\al_e^{-1}(M_Z) = \ {128}\,, \qquad
\sin^2 \theta_W(M_z)=\  0.231 \,,\qquad  
M_Z = \ 91.2 \text{ GeV} \,, \qquad
M_W =\  80.4 \text{ GeV}  \,,
\ee
and choose $M_W$ as the argument of the Sudakov logarithm.
The high-energy asymptotic of the one-loop corrections read
\begin{align}
	\dd \si_2^{(1)} = &-0.34+1.10x-1.66x^2+1.10x^3-0.34x^4 \,,\\[3mm]
	\dd \si_1^{(1)} =&\, 1.21-3.33x+4.99x^2-3.33x^3+1.21x^4 \nonumber \\
		&\quad+ \left(0.10-0.52x+0.78x^2-0.52x^3+0.10x^4\right)\ln(1-x) \nonumber \\
		&\quad- \left(0.02-0.49x+2.05x^2-2.24x^3+0.75x^4\right)\ln(x) \,,\\[3mm]
	\dd \si_0^{(1)} =& -0.49+1.78x+2.60x^2-5.34x^3+2.61x^4 \nonumber \\
		&\quad+0.03\left(x-x^2+x^3\right)\ln^2(1-x) \nonumber \\
		&\quad -\left(0.05-0.16x+0.28x^2-0.16x^3+0.05x^4\right)\ln(1-x) \nonumber \\
		&\quad +\left(0.31-0.24x-0.60x^2+0.48x^3\right)\ln^2(x) \nonumber \\
		&\quad +\left(1.21-2.51x+2.64x^2-0.97x^3+0.05x^4\right)\ln(x) \nonumber \\
		&\quad +\left(0.10-0.44x+0.42x^2-0.13x^3\right)\ln(1-x)\ln(x)  \,.
\end{align}

\begin{figure}[t]
\begin{center}
\begin{tabular}{cc}
\includegraphics[height=1.9in]{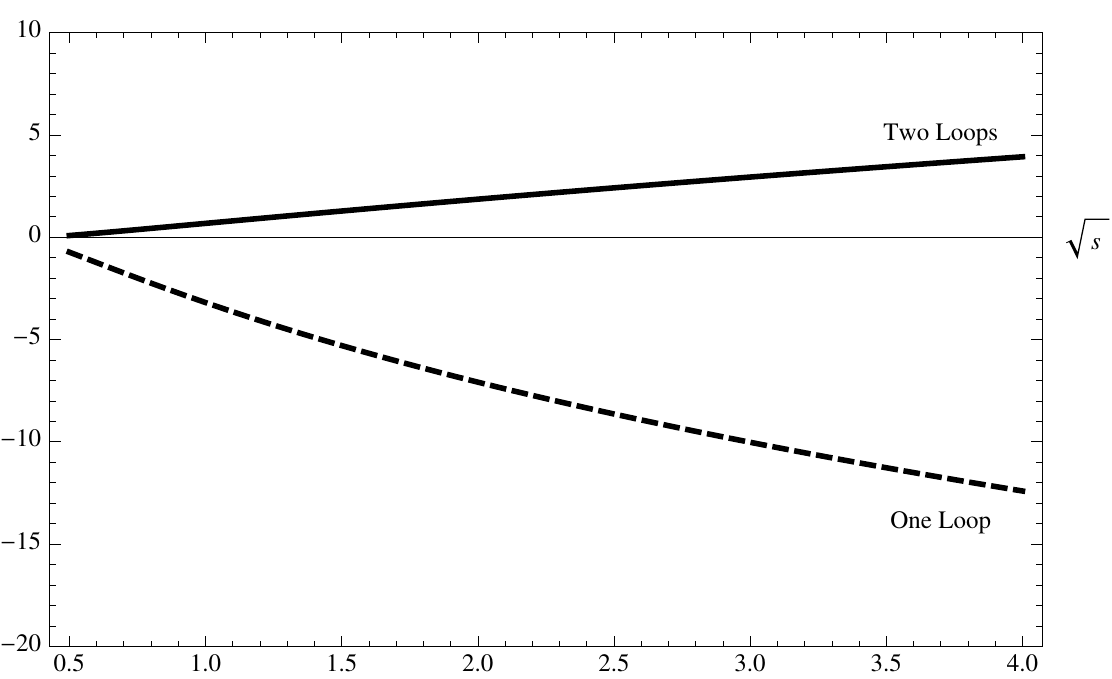} & \includegraphics[height=1.9in]{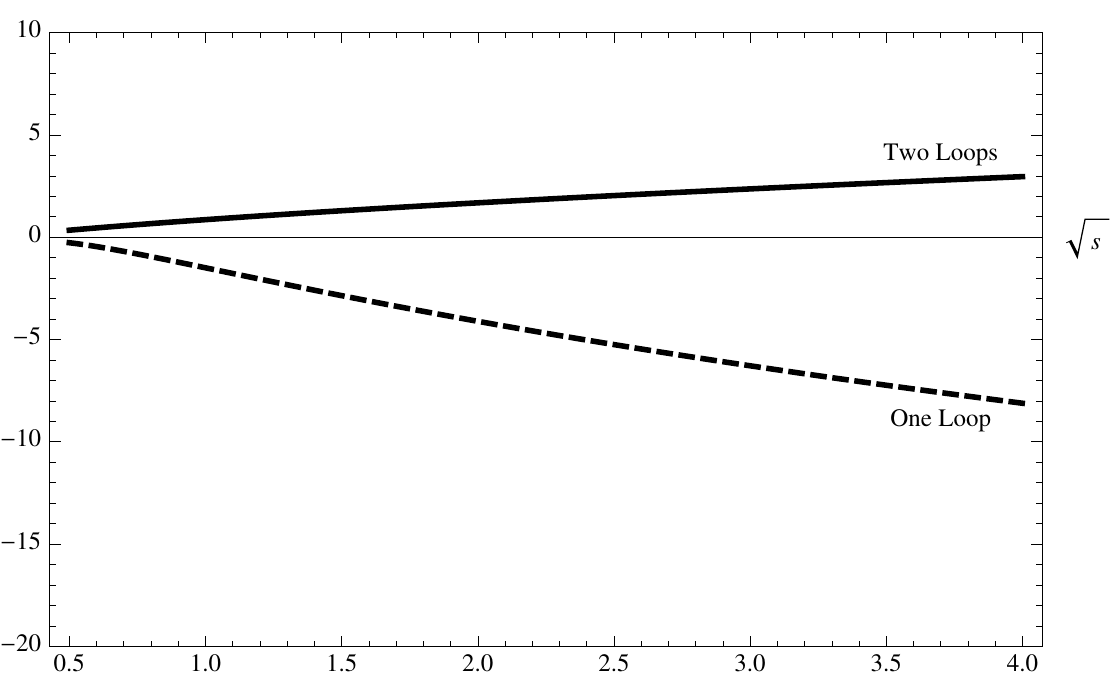} \\
(a) $\theta=50^\circ$ & (b)  $\theta=90^\circ$
\end{tabular}
\end{center}
\caption{\label{fig2}
The one- and two-loop electroweak corrections in \% to the Born cross section
as function of the center-of-mass energy for (a) $\theta=50^\circ$ and (b) $\theta=90^\circ$.} 
\end{figure}

The two-loop LL, NLL, and NNLL  electroweak corrections are given by
Eqs.~(\ref{tlewnn}) in the Appendix B. The   N$^3$LL electroweak correction  is
sensitive to fine details of the gauge boson mass generation  and currently the
exact result is not available. We evaluate them approximately  within a  pure
$SU_L(2)$ model considered in the previous section.  The numerical result is
given by Eq.~(\ref{tlewnnn}) in the Appendix B. Note that Eq.~(\ref{tlewnnn})
correspond to the decomposition~(\ref{ewseries}) where all the coupling constants
are normalized at the gauge boson mass ({\it cf.} Eq.~\ref{su2series}). In the
full standard model the hypercharge interaction and gauge boson mixing result in
a linear-logarithmic contribution which is not accounted for in this
approximation.  It is, however, suppressed by the small factor
$\sin^2\theta_W\approx 0.2$.  Therefore, the approximation gives an estimate of
the coefficient in front of the linear electroweak logarithm with $20\%$
accuracy.  As we will see, such an uncertainty in the coefficient of the
two-loop linear logarithm is  negligible for practical applications. 

\begin{figure}[t]
\begin{center}
\begin{tabular}{cc}
\includegraphics[height=1.9in]{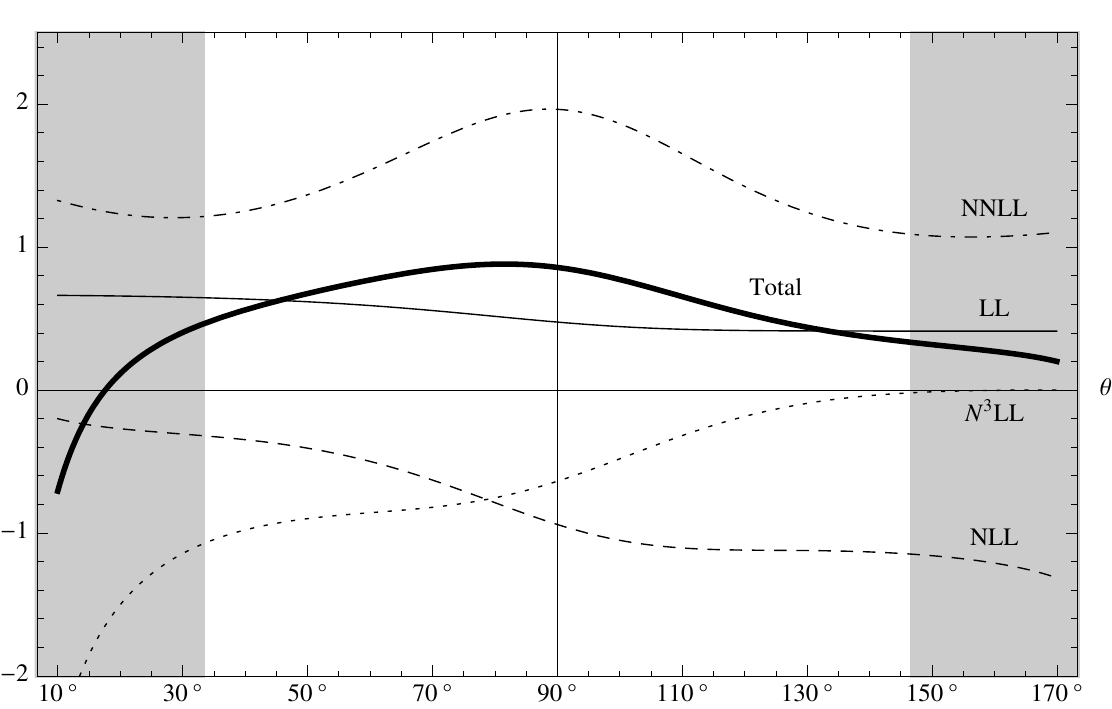} & \includegraphics[height=1.9in]{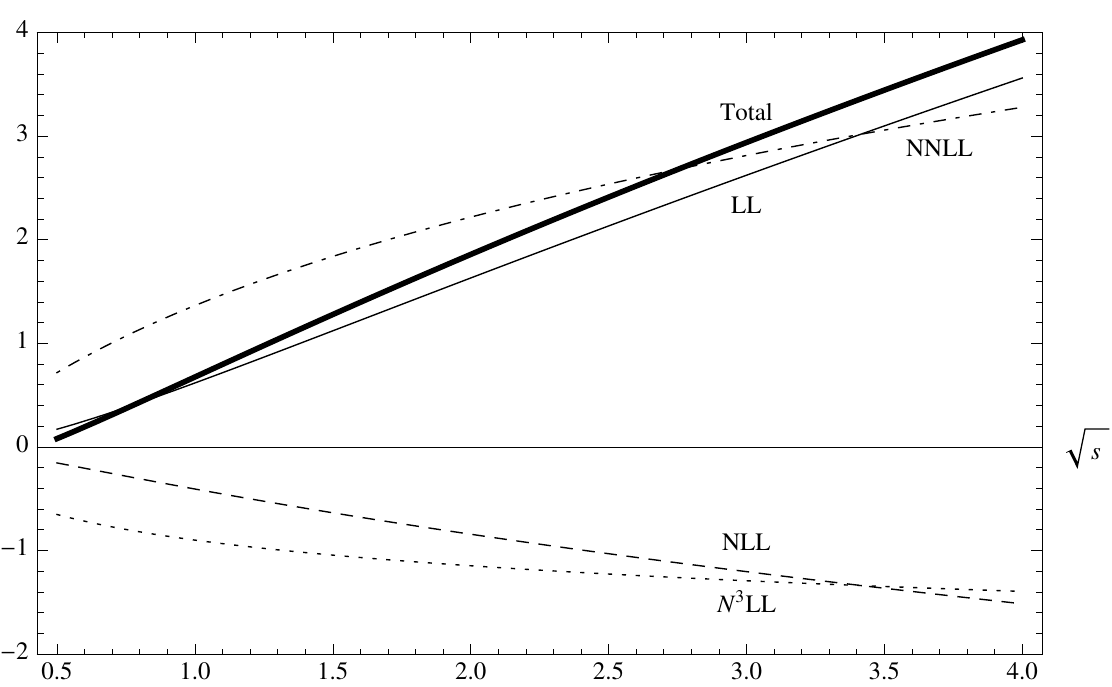} \\
(a) $\sqrt{s} = 1$~TeV & (b)  $\theta=50^\circ$
\end{tabular}
\end{center}
\caption{\label{fig3}   
Separate two-loop logarithmic contributions in \% to the Born cross section
as function of  (a) the scattering angle  for $\sqrt{s} =1$~TeV
and (b) the center-of-mass energy for  $\theta=50^\circ$. The shaded 
area corresponds to the region where Sudakov approximation is not relaible.} 
\end{figure}

Let us now consider the numerical effect of the corrections. We plot the one-
and two-loop electroweak corrections relative to the Born cross section for the
phenomenologically interesting intervals of the center-of-mass energy and the
scattering angle in Fig.~\ref{fig1} and \ref{fig2}. For the small and very large scattering
angles the Sudakov approximation is not reliable since there either the $t$ or $u$
invariant is comparable with $M^2_{W,Z}$ and the power suppressed terms become
important.  In Fig.~\ref{fig1} the shaded regions denote the kinematic
domain where either $t$ or $u$ are less than $10 M_Z^2$ and the Sudakov limit
used in this calculation begins to falter.  The one-loop correction strongly
depends on the center-of-mass energy and scattering angle and is large and
negative in the practically interesting region. Overall the two-loop correction is
positive and enters at the 1 percent level everywhere in the angular
distribution, with the largest corrections coming at $\theta \sim 90^\circ$. The
breakdown of the two-loop corrections into the individual logarithmic
contributions is given in Fig.~\ref{fig3}. The corrections exhibit typical 
sign-alternating behavior with significant cancellation of the individual
logarithmic terms.  For example,  at $\sqrt{s} = 1$ TeV and $\theta = 50^\circ$
the numerical structure of the corrections read  
\be
\text{One loop: } -3.20\% \left \{ \begin{array}{cc}
\text{LL: } & -9.45 \% \\
\text{NLL: }& 7.35\% \\
\text{N$^2$LL: } & -1.09\% \\
\end{array} \right .  \qquad \text{Two loops: } 0.67\%\quad \left \{ \begin{array}{cc}
\text{LL: } &0.62\% \\
\text{NLL: }&-0.41\% \\
\text{N$^2$LL: }&  1.36\% \\
\text{N$^3$LL: }& -0.90\%
\end{array} \right .  \,
\ee
Thus a twenty percent uncertainty in the coefficient of the two-loop linear
logarithm would only induce  an uncertainty  on the order of one permill in the cross
section. 

To conclude, we have computed the dominant two-loop electroweak corrections to
high-energy wide-angle Bhabha scattering.  The corrections can be as large as
10\% in one loop and 1\% in two loops. Our  result completes the perturbative
analysis of the Bhabha scattering necessary for the luminosity determination at
the ILC.

\acknowledgments
The work is supported in part by NSERC and  Alberta Ingenuity foundation. 
The work of A.P. is also supported by DFG Mercator grant.

\newpage

\appendix

\section{Two loop results in $SU_L(2)$ model}
Here we present the analytical result for the two-loop logarithmic
corrections to the cross section of Bhabha scattering in the $SU_L(2)$ model
with the gauge and the Higgs boson of the same mass $M$. The annihilation
contribution reads 
\begin{align}
	\expn{\dd \si_{S\ 4}}{2} &=  \frac{9}{2} \ , \nonumber \\
	\expn{\dd \si_{S\ 3}}{2} &= -\frac{143}{6}-6\ln(1-x) + 30\ln(x) \ , \nonumber \\
	\expn{\dd \si_{S\ 2}}{2} &= -\frac{145}{4}+11\pi^2+\frac{89}{6}\ln(1-x)+2\ln^2(1-x) - 8\ln(x)\ln(1-x)  \nonumber \\
		&\qquad-\frac{(535-445x)\ln(x)}{6(1-x)} + \frac{(91-182x+76x^2)\ln^2(x)}{2(1-x)^2} \ , \nonumber \\
	\expn{\dd \si_{S\ 1}}{2} &= \frac{28411}{216}-122 \zeta(3)+26\sqrt{3}\text{Cl}_2\! \left(\frac{\pi}{3}\right) + 15\sqrt{3}\pi - \frac{(199-127x)\pi^2}{6(1-x)} \nonumber \\
		&\qquad +\left(\frac{125}{3}+\frac{50\pi^2}{3}+\frac{10\ln(x)}{1-x}-\frac{5(1-2x)\ln^2(x)}{(1-x)^2}\right)\ln(1-x) \nonumber \\
		&\qquad +\left(-\frac{1075-1250x}{6(1-x)}+\frac{(74-148x+38x^2)\pi^2}{3(1-x)^2}\right)\ln(x) \nonumber \\
		&\qquad - \frac{(631-806x)\ln^2(x)}{4(1-x)^2} + \frac{19(1-2x)\ln^3(x)}{(1-x)^2} \,.
\label{tlsus}
\hspace*{50mm}
\end{align}
The scattering contribution reads 
\begin{align}
	\expn{\dd \si_{T\ 4}}{2} &=  \frac{9}{2} \ , \nonumber \\
	\expn{\dd \si_{T\ 3}}{2} &= -\frac{143}{6}-6\ln(1-x) - 6\ln(x) \ ,\nonumber \\
	\expn{\dd \si_{T\ 2}}{2} &= -\frac{145}{4}+8\pi^2+\frac{89}{6}\ln(1-x)+2\ln^2(1-x) - 14\ln(x)\ln(1-x)  \nonumber \\
		&\qquad+\frac{(41-131x)\ln(x)}{6(1-x)} - \frac{(26-22x+11x^2)\ln^2(x)}{2(1-x)^2}  \ , \nonumber \\
	\expn{\dd \si_{T\ 1}}{2} &= \frac{28411}{216}-122\zeta(3)+26\sqrt{3}\text{Cl}_2\! \left(\frac{\pi}{3}\right) + 15\sqrt{3}\pi + \frac{(8+4x)\pi^2}{1-x} \nonumber \\
		&\qquad +\left(\frac{125}{3}+\frac{8\pi^2}{3}+\frac{(17-7x)\ln(x)}{1-x}-\frac{(10-30x+15x^2)\ln^2(x)}{(1-x)^2}\right)\ln(1-x) \nonumber \\
		&\qquad +4\ln(x)\ln^2(1-x)+\left(\frac{1030-505x}{18(1-x)}+\frac{(80-88x+44x^2)\pi^2}{3(1-x)^2}\right)\ln(x) \nonumber \\
		&\qquad + \frac{(414-502x+263x^2)\ln^2(x)}{12(1-x)^2} + \frac{(10-22x+11x^2)\ln^3(x)}{(1-x)^2} \,,
\label{tlsut}
\hspace*{25mm}
\end{align}
The interference contribution reads 
\begin{align}
	\expn{\dd \si_{ST\ 4}}{2} &=  \frac{9}{2}  \ , \nonumber \\
	\expn{\dd \si_{ST\ 3}}{2} &= -\frac{143}{6}-6\ln(1-x) +12\ln(x) \ , \nonumber \\
	\expn{\dd \si_{ST\ 2}}{2} &= -\frac{145}{4}-\frac{17}{2}\pi^2+\frac{89}{6}\ln(1-x)+2\ln^2(1-x) - 11\ln(x)\ln(1-x)  \nonumber \\
		&\qquad-\frac{(247-157x)\ln(x)}{6(1-x)} - \frac{(7+16x+7x^2)\ln^2(x)}{4(1-x)^2} \ , \nonumber \\
	\expn{\dd \si_{ST\ 1}}{2} &= \frac{28411}{216}-122\zeta(3)+26\sqrt{3}\text{Cl}_2\! \left(\frac{\pi}{3}\right) + 15\sqrt{3}\pi - \frac{67\pi^2}{12} \nonumber \\
		&\qquad +\left(\frac{125}{3}+\frac{11\pi^2}{3}+\frac{(27-7x)\ln(x)}{2(1-x)}-\frac{(27-64x+27x^2)\ln^2(x)}{2(1-x)^2}\right)\ln(1-x) \nonumber \\
		&\qquad +2\ln(x)\ln^2(1-x)-\left(\frac{2195-3245x}{36(1-x)}+\frac{(13+41x)\pi^2}{3(1-x)}\right)\ln(x) \nonumber \\
		&\qquad - \frac{(49+32x-431x^2)\ln^2(x)}{24(1-x)^2} - \frac{(8+2x^2)\ln^3(x)}{(1-x)^2} \,.
\label{tlsust}
\hspace*{38mm}
\end{align}
For the $e^+e^-\to\nu\bar{\nu}$ annihilation cross section the two-loop
logarithmic corrections read
\begin{align}
	\expn{\dd \si_{S\ 4}}{2} &=  \frac{9}{2} \ , \nonumber \\
	\expn{\dd \si_{S\ 3}}{2} &=  -\frac{143}{6}+30\ln(1-x) -6\ln (x) \ , \nonumber \\
	\expn{\dd \si_{S\ 2}}{2} &=  -\frac{145}{4}+11\pi^2 -\frac{445}{6}\ln(1-x) +38\ln^2(1-x)-8\ln(x)\ln(1-x) \nonumber \\
		& \qquad  +\frac{(107-89x)\ln(x)}{6(1-x)}+\frac{(1-2x+4x^2)\ln^2(x)}{2(1-x)^2}  \ , \nonumber \\
	\expn{\dd \si_{S\ 1}}{2} &= \frac{28411}{216}-122 \zeta(3)+26\sqrt{3}\text{Cl}_2\! \left(\frac{\pi}{3}\right) + 15\sqrt{3}\pi - \frac{(55-127x)\pi^2}{6(1-x)} \nonumber \\
		&\qquad +\left(-\frac{625}{3}+\frac{38\pi^2}{3}-\frac{2\ln(x)}{1-x}+\frac{(1-2x)\ln^2(x)}{(1-x)^2}\right)\ln(1-x) \nonumber \\
		&\qquad +\left(\frac{215-250x}{6(1-x)}+\frac{(14-28x+50x^2)\pi^2}{3(1-x)^2}\right)\ln(x) \nonumber \\
		&\qquad + \frac{(11-46x)\ln^2(x)}{4(1-x)^2} + \frac{(1-2x)\ln^3(x)}{(1-x)^2} \,.
\label{tlsusnu}
\hspace*{50mm}
\end{align}

\section{Two loop electroweak numerical results}
The two-loop electroweak logarithmic corrections to the Bhabha cross section to the NNLL approximation read
\begin{align}
	\dd \si_4^{(2)} = &\ 0.34-1.21x+1.81x^2-1.21x^3+0.34x^4 \,,\nonumber\\[2mm]
	\dd \si_3^{(2)} =& -1.43+3.51x-4.81x^2+2.91x^3-1.07x^4 \nonumber \\
		&\quad -\left(0.16-0.81x+1.22x^2-0.81x^3+0.16x^4\right)\ln(1-x) \nonumber \\
		&\quad - \left(0.18+0.74x-4.23x^2+4.91x^3-1.71x^4\right)\ln(x) \,,\nonumber\\[2mm]
	\dd \si_2^{(2)} =& \ 5.78 + 6.42 x - 20.04 x^2 + 8.19 x^3 + 3.98 x^4 \nonumber \\
		&\quad +\left(0.05 - 0.21 x + 0.30 x^2 - 0.21 x^3 + 0.05 x^4\right)\ln^2(1-x)\nonumber \\
		&\quad +\left(0.97 - 3.27 x + 3.62 x^2 - 1.48 x^3 - 0.13 x^4\right) \ln(1-x) \nonumber \\
		&\quad -\left(0.29 + 0.29 x - 4.06 x^2 + 5.22 x^3 - 1.97 x^4\right) \ln^2(x) \nonumber \\
		&\quad -\left(3.74 - 7.56 x + 9.03 x^2 - 4.45 x^3 + 0.43 x^4\right) \ln(x) \nonumber \\
		&\quad -\left(0.88 - 2.76 x + 2.13 x^2 - 0.10 x^3 - 0.28x^4\right) \ln(1-x) \ln(x)  \,.
\label{tlewnn}
\end{align}
The numerical result for the two-loop N$^3$LL corrections to the  Bhabha cross section in the $SU_L(2)$ 
model reads
\begin{align}
	\dd \si_1^{(2)} =& -0.30+0.87x+0.41x^2-2.21x^3+1.24x^4 \nonumber \\
	&\quad +\left(4.25-18.23x+36.60x^2-35.50x^3+12.89x^4\right)\ln(1-x) \nonumber \\
	&\quad +\left(0.63-0.38x+1.88x^2-2.13x^3\right)\ln^3(x) \nonumber \\
	&\quad +\left(0.18-0.38x-1.75x^2-0.03x^3\right)\ln^2(x) \nonumber \\
	&\quad +\left(18.97-10.32x+6.39x^2-9.83x^3-5.21x^4\right)\ln(x) \nonumber \\
	&\quad +\left(0.25-0.75x+0.75x^2-0.25x^3\right)\ln^2(1-x)\ln(x) \nonumber \\
	&\quad -\left(0.63-3.56x+5.25x^2-2.31x^3\right)\ln(1-x)\ln^2(x) \nonumber \\
	&\quad +\left(0.27-1.60x+2.40x^2-1.06x^3\right)\ln(1-x)\ln(x) \,.
\label{tlewnnn}
\end{align}


\end{document}